\documentclass{egpubl}
 
 \JournalSubmission    
\CGFccby

\usepackage[T1]{fontenc}
\usepackage{dfadobe}  

\usepackage{cite}  
\BibtexOrBiblatex
\electronicVersion
\PrintedOrElectronic
\ifpdf \usepackage[pdftex]{graphicx} \pdfcompresslevel=9
\else \usepackage[dvips]{graphicx} \fi

\usepackage{egweblnk} 

\title[\framework]%
      {\framework: Multiview Augmented Reality Visualization\\ for Exploring Rich Material Data}

\author[Alexander Gall]
{\parbox{\textwidth}{\centering A. Gall\thanks{e-mail: alexander.gall@uni-passau.de}$^{1,2}$\orcid{0000-0002-3649-7368}
        and A. Heim$^{1,3}$\orcid{0000-0002-3670-5403} 
        and E. Gr{\"o}ller$^{1}$\orcid{0000-0002-8569-4149} 
        and C. Heinzl$^{2,3}$\orcid{0000-0002-3173-8871} 
        }
        \\
{\parbox{\textwidth}{\centering 
        $^1$ TU Wien, Faculty of Informatics \\
        $^2$ University of Passau, Faculty of Computer Science and Mathematics\\
        $^3$ Fraunhofer Institute for Integrated Circuits IIS, Division Development Center X-ray Technology\\
       }
}
}

\usepackage{siunitx} 
\usepackage{xspace} 
\xspaceaddexceptions{]\}}

\newcommand{\framework}{MARV\xspace}
\newcommand{\firstVis}{MDD Glyphs\xspace}
\newcommand{\secondVis}{Temporal Evolution Tracker\xspace}
\newcommand{\secondVisAcro}{TET\xspace}
\newcommand{\thirdVis}{Chrono Bins\xspace}
\newcommand{\colorWidget}{Skewness Kurtosis Mapper\xspace}
\newcommand{\colorWidgetAcro}{SK Mapper\xspace}

\usepackage{xcolor}
\definecolor{newcolor}{rgb}{.8,.349,.1}

\begin{document}

 \teaser{
  \includegraphics[width=0.81\linewidth]{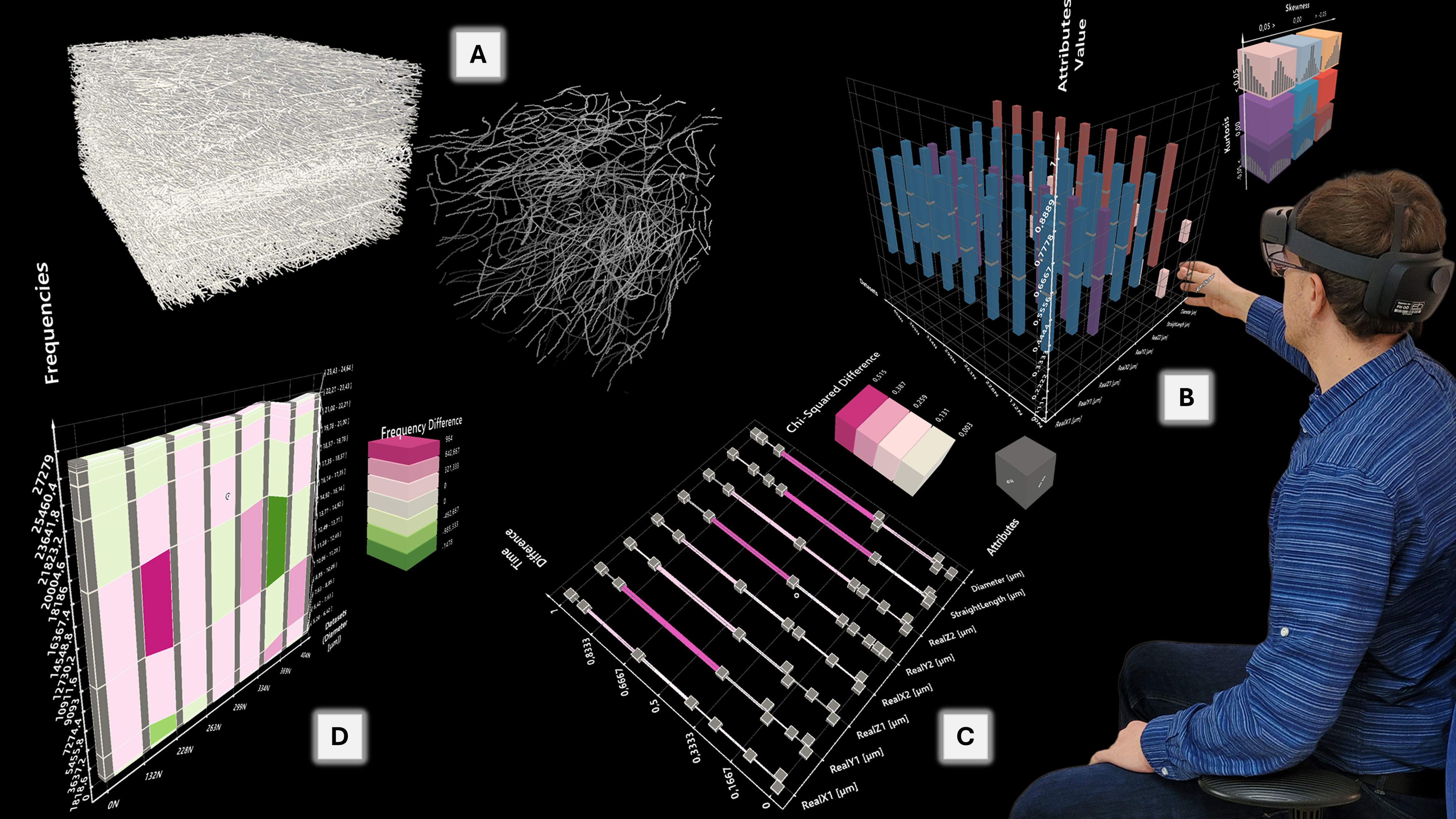}
  \centering
   \caption{
   The \framework workspace: \framework facilitates tailored analyses of rich material data in a seamless immersive experience, exploring spatial data through 3D renderings (A), as well as abstract data through \firstVis and \colorWidget (B), \secondVis (C), and \thirdVis (D).} 
 \label{fig:teaser}
}

\maketitle
\begin{abstract}
   Rich material data is complex, large and heterogeneous, integrating primary and secondary non-destructive testing data for spatial, spatio-temporal, as well as high-dimensional data analyses. Currently, materials experts mainly rely on conventional desktop-based systems using 2D visualization techniques, which render respective analyses a time-consuming and mentally demanding challenge. 
   \framework is a novel immersive visual analytics system, which makes analyses of such data more effective and engaging in an augmented reality setting. For this purpose \framework includes three newly designed visualization techniques: \firstVis with a \colorWidget, \secondVis, and \thirdVis, facilitating interactive exploration and comparison of multidimensional distributions of attribute data from multiple time steps. A qualitative evaluation conducted with materials experts in a real-world case study demonstrates the benefits of the proposed visualization techniques. This evaluation revealed that combining spatial and abstract data in an immersive environment improves their analytical capabilities and facilitated the identification of patterns, anomalies, as well as changes over time.


\end{abstract}  

\section{Introduction} 

    
    Composite materials and especially fiber-reinforced polymers (FRPs) are used in many industrial applications such as automotive, or aerospace, due to their high strength-to-weight ratio and resistance to many external influences~\cite{Heinzl2017}. The properties of FRPs result from the combination of different microstructural components (e.g., fibers, matrix) and their respective characteristics (e.g., fiber length, diameter). X-ray computed tomography (XCT) as a non-destructive testing (NDT) method enables detailed characterizations of material systems through an assessment of the internal microstructure \cite{Ida2019, Heinzl2019, Rajak_2019}. XCT generates primary data in the form of 3D volumetric datasets, which can be further enriched by the derivation of secondary data, quantifying microstructural features of interest.
    Comparisons between material systems are necessary to develop materials to meet specific requirements. Experts need to understand spatial \emph{and} quantitative differences in rich material data (i.e., primary and secondary data). In the case of FRPs these consist of XCT scans including hundreds of thousands of fibers, each quantified by more than 25 different attributes. This highly complex and cognitively demanding task becomes even more challenging if materials require comparisons throughout a longer period of time, as in the case of in-situ loading experiments. 
    
    Traditional 2D visualization techniques are unsuitable for the preservation of critical spatial relationships and structural details within XCT datasets, which are inherently three-dimensional \cite{Dwyer2018}. Recent research indicates that the exploration of complex spatial data can be improved using immersive analysis techniques \cite{Billinghurst2018, Kraus2021, Kraus2022, Ens2021a}. Augmented Reality (AR) is increasingly adopted in industrial applications due to its ability to seamlessly integrate digital information into real-world environments, enabling intuitive, hands-free interaction and enhancing collaboration \cite{Masood2020, Evangelista2020}. This makes AR particularly valuable in scenarios that require flexible on-site inspection, such as material characterization and quality control, where it supports quick, and efficient analysis \cite{Zhou2018, Ho2022}.
    
    In this work our goal is to enable domain experts to explore materials as well as their properties in an intuitive, more efficient and engaging way by enabling depth perception, efficient 3D navigation, and detailed analysis of time-varying in-situ tests. For this purpose, we have developed a generalizable framework called \textbf{M}ultiview \textbf{A}ugmented \textbf{R}eality \textbf{V}isualization (\framework), designed to support complex spatial representations and high-dimensional abstract data visualization.
    
    Advancing the state of the art in terms of immersive analytics (IA) of rich material data, our core contributions are:
    \begin{itemize}
        \item A novel augmented reality framework for analyzing spatial and multi-dimensional abstract data, and its application to a real world problem from the domain of materials science.
        \item Introduction of three novel immersive visualization techniques, developed in close collaboration with materials experts:
        \begin{itemize}
            \item Multidimensional Distribution Glyphs (\firstVis) and the \colorWidget provide a visual summary of statistical characteristics of the attributes of interest. 
            \item \secondVis visualizes changes in the distributions of these attributes over time. 
            \item \thirdVis allow to compare the severity of changes between datasets.
        \end{itemize}
        \item A qualitative evaluation based on a real-world case study with material specialists and novices. The preliminary evaluation serves as basis for assessing the viability of our proposed AR framework and the developed visualization techniques.
    \end{itemize}

\section{Related Work} \label{sec:RelatedWork} 
    Conventional visual analysis in materials science mainly involves 2D desktop monitor based setups using mouse and keyboard as input devices. Currently, there are only a few solutions that support material experts with IA methods and enable them to use spatial immersion and embodied navigation, leveraging their own spatial cognition skills. In a recent systematic review, Saffo et al.~\cite{Saffo2023} identified three immersive applications between 2013 and 2022 featuring both volumetric and tabular data. Therefore, we also present studies from approaches dealing with either spatial or abstract data. These cases are necessary to help identify solutions for our real-world problem, and highlight the gaps still present in the research, particularly in AR, which we address in our work.
      
      
\subsection{Visual Analytics}
    Materials experts usually rely on traditional desktop-based visual analysis techniques integrating 2D cross-sectional views and 3D volume renderings, but also 2D charts and tabular overviews, to accurately evaluate and characterize multidimensional material data. Open source tools such as Paraview~\cite{ParaView}, Quanfima~\cite{Shkarin_2019}, and open\_iA~\cite{Froehler_2019} offer visual analysis on 2D monitors to experts. Furthermore, proprietary software tools such as VGStudio MAX~\cite{Volumegraphics} or Avizo~\cite{ThermoFisher} are commonly used to analyze the properties of volumetric data of complex material systems. All of these systems share the disadvantage that they cannot facilitate the analysis of inherently spatial features in their native domain. When localizing defects in the volume or tracking curved fibers, depth perception is lacking and navigation within or around 3D objects is difficult. In addition, these systems can only be analyzed at a stationary workstation, which requires multiple monitors to allow experts the comparison of time-varying datasets.
    
\subsection{Immersive Analytics}
    IA offers promising opportunities in various application areas \cite{Fonnet2019, Ens2021a, Saffo2023, Froehler2022}. Particularly relevant to the presented materials science use case is the increase in productivity, accuracy, and autonomy in the quality sector \cite{Evangelista2020, Ho2022} made possible by AR as the key technology.
    An overview of the most relevant immersive applications is presented below, accompanied by a more detailed comparison against \framework in the supplemental material.

    A notable application within the domain of \textbf{virtual reality (VR)} to visualize large multivariate \textbf{abstract data} is DataHop \cite{Hayatpur2020}. It allows to spatially layout data analysis workflows, based on different charts, and spatially traverse modifications in the data. Reski et al.~\cite{Reski2020} proposed a 3D radar chart to facilitate spatial interaction, annotation, and analysis of time-oriented data. Wagner et al.~\cite{Wagner2024} introduced an immersive space-time cube metaphor for urban data exploration, enabling spatio-temporal queries and bi-manual interactions.
    These frameworks demonstrate the positive impact of immersive data exploration but do not support the analysis of rich material data in an AR workspace. For material characterization it is essential to visualize primary data and link them to their associated secondary data.

    Several studies have analyzed \textbf{abstract data in AR} systems. The IATK~\cite{Cordeil2019} and DXR~\cite{Sicat2019} frameworks enable visualization of large multivariate data in VR and AR. They support rapid prototyping through a declarative visualization grammar and reusable chart templates utilizing the desktop environment. Sardana et al.~\cite{Sardana2021} compared spatio-temporal data visualization with AR head mounted displays (HMDs) with conventional 2D displays. Hubenschmid et al.~\cite{Hubenschmid2021} demonstrated STREAM, a system that enables fluid interaction between multimodal 2D and 3D visualizations through AR HMDs, room-tracked tablets, and a large screen. The proposed system of Satriadi et al.~\cite{Satriadi2022a} explores linking physical objects with multivariate charts for data visualization. In contrast, for all of our material analysis tasks the visualization of spatial information in combination with its abstract data is crucial. Furthermore, Satkowski et al.~\cite{Satkowski2021} investigated the effects of the real environment on the perception of 2D visualizations and concluded that background objects have only a minor influence on the measured performance. 
    Although the aforementioned studies do not provide systems for exploring volume renderings or for analyzing time-varying rich material data, they do demonstrate the opportunities that an AR system can provide.

    There are only a few approaches using \textbf{VR} HMDs, focusing on the visualization of data with an \textbf{inherent spatial structure}. FiberClay~\cite{Hurter2019} is an immersive platform for visualizing and comparing 3D aircraft trajectories and human brain fiber traces. The system employs 3D selections and smooth transitions between mappings, outperforming 2D systems in terms of pattern recognition.
    Considering FRPs, their system falls short in terms of visualizing all dimensions of the secondary data. AeroVR~\cite{Tadeja2020} addresses the design process of aerospace components through immersive renderings of computer aided design (CAD) models with 3D scatter plots. It is not suitable for material characterization tasks as it does not allow an analysis of XCT scans or time-varying data. Wang et al.~\cite{Wang_2020} are concerned with the direct manipulation of XCT datasets and transfer functions using hand gestures, lacking the ability to visualize and analyze additional multi-dimensional data. Gall et al.~\cite{Gall2021} developed the ImNDT immersive analytics system for multidimensional secondary data exploration of FRPs in VR, which provided advantages in structural pattern detection. ImNDT was extended for Cross-Virtuality-Analysis introducing a mixed reality mode for collaborative analysis via a large touchscreen \cite{Gall2021b}.
    While both tools were beneficial, they do not have the capability to render XCT volumes or compare time-varying material data.
    
    An early study using \textbf{AR} to \textbf{visualize spatial data} was presented by Schickert et al.~\cite{Schickert2018}, who visualized NDT and CAD data on physical objects using smartphones. Ferraguti et al.~\cite{Ferraguti2019} demonstrated an AR HMD approach for quality assessment of polished surfaces by projecting metrology data onto the specimen. The medical application of Pratt et al.~\cite{Pratt_2018} uses AR to superimpose 3D vascular models onto the patient, enhancing precise localization of perforated vessels. VoxAR by Boorboor et al.~\cite{Boorboor2023} aims to improve the volume visualization of see-through HMDs by adjusting the position and coloring based on the real-world background. 
    Previous systems lack embodied interaction and advanced multidimensional data visualization, which are essential for complex material inspection. 
    Embodied Axes~\cite{Cordeil2020} enhances AR selections in XCT scans and data visualization with tangible controllers. 
    The system's inability to explore high-dimensional abstract data with spatial information and time-varying data necessitates further research.
    
                

\section{Background \& Design Principles} \label{sec:Background}  

    Since our study is a problem-oriented research, we apply the design study methodology proposed by Sedlmair et al. \cite{Sedlmair2012}. In this section, we provide a comprehensive overview of the visualization tasks and research questions, as well as our design considerations for the final setup of the immersive analytics system \framework.

    \subsection{Preliminary Study}
        To the best of the authors' knowledge, there is currently no AR analysis framework for interpreting spatial and abstract data from materials science in a single comprehensive tool. So, it is crucial to identify the user groups and their specific tasks, as well as to determine the most effective design for supporting their work. The \textit{Learn} stage of the design study methodology \cite{Sedlmair2012} is represented in \autoref{sec:RelatedWork}, which provides a foundation for understanding current problems and solutions in our domain and informs the initial stages of our study. 
        In the course of the \textit{Winnow} and \textit{Cast} stage, two domain experts from educational institutions with extensive experience in materials characterization were identified. Both have been working with rich material data on a daily basis for an average of 15 years. Over a period of six months, we engaged in close collaboration with these experts, conducting a series of semi-structured interviews and observations of their workflows in order to identify their needs and challenges.\\
        Regarding our goal and task analysis, we applied the \textit{Where-What-Who-Why-How} design framework by Marriott et al.~\cite{Marriott2018b}. It represents an extension of the \textit{What-Why-How} framework by Brehmer and Munzner~\cite{Brehmer2013}, specifically adapting it for immersive analytics and aligning well with the \textit{Discover} and \textit{Design} stages of Sedlmair's design study methodology.

        \noindent\textbf{Where:} Our system should support experts both at desktops and outside their offices. Addressing this requirement, integrated see-trough AR HMDs, such as Microsoft HoloLens 2 (MH2), are currently the most promising solution. Such HMDs facilitate maintaining visibility of industrial environments without video delay and support hands-free control, enabling interaction with physical objects present in the surrounding. Inside-out tracking of these AR devices additionally allows for unobstructed collaboration with co-workers. Finally, it is possible to use complementary interfaces \cite{Zagermann2022} with desktop applications.\\ 
        %
        \textbf{What:} A vital method to investigate damage mechanisms in material systems is in-situ tensile testing. Such experiments involve sequential scans of a specimen under increasing load, yielding a sequence of primary datasets, i.e., time steps. Secondary information is extracted via the fiber characterization pipeline of Salaberger et al.~\cite{Salaberger2011} and consists of several hundred thousands features of interest (i.e., fibers), with more than 25 attributes each. Additional statistical data (e.g., kurtosis, skewness) is computed and utilized in the examination of the material data. For each step in the experiment, the data thus contains both volumetric primary data and multidimensional secondary data extracted from the features of interest. Our framework aims to ensure generalizability across data from different domains, but focuses in this study on materials science data.\\ 
        \textbf{Who:} This study primarily focuses on rich material data, making both experts and novices in the field of materials science our target users. Experts are versed in the analysis of large amounts of rich material data using 2D charts and slice views on multiple office monitors. For these experts, the focus lies on analyzing and comparing temporal changes. Furthermore they investigate distributions of attributes describing material characteristics to identify anomalies and trends efficiently. Novices, on the other hand, may struggle with envisioning the 3D structure based on numeric data alone, making immersive 3D visualization and the ability to highlight changes in spatial representations critical to their understanding.\\
        \textbf{Why:} Our goal is to offer a compact overview of rich time-varying material data enabling users to conduct a standalone analysis. We identified the high-level user task of discovery, which involves an exploratory search for patterns and changes in the abstract data and in the spatial volume \cite{Brehmer2013}. Visualization tasks, such as characterizing distributions, finding anomalies, identifying correlations/trends, making comparisons \cite{Lee2017}, filtering, and sorting \cite{Saket2019} are especially critical for material experts. The search is driven by exploration, so both target and location are unknown. Once a target or set of targets has been found, compare or summarize tasks are possible to setup new hypotheses, along with verifying or falsifying existing ones. The difficulty of these tasks lies in the fact that both spatial representations and abstract information are required, as well as the comparison of temporal changes in order to enable a material characterization.\\
        \textbf{How:} The targeted framework needs to enable users to observe primary XCT data through Direct Volume Rendering (DVR) and to explore secondary data using abstract data visualization techniques. Using gestures, both spatial and abstract data can be manipulated or filtered, resulting in an engaging, interactive and comprehensive data analysis experience. In contrast to the current state of the art, \framework offers a unified framework that combines both primary and secondary material data with immersive analytic techniques. Furthermore, it provides novel visualization techniques and an innovative view-dependent interaction that enable experts to conduct comprehensive inspections of in-situ tests. 

    \subsection{Goals, Tasks \& Requirements} 
    \label{subsec:Requirements} 

        %
        Based on the insights we gained during the discovery and design phases, we identified several challenges for the analysis of material data requiring depth perception and 3D navigation. Current immersive systems do not adequately support the analysis of rich material data in a unified, intuitive environment, which makes a comprehensive and detailed inspection challenging for experts. Furthermore, comparing a large number of scans from in-situ tests across multiple time steps is a time-consuming and error-prone process for domain experts. Existing approaches have limitations in effectively visualizing shifts in the distribution of material properties in a holistic view and for detailed changes of features and characteristics.
        Based on these challenges, identified in the preliminary study, we have defined a set of domain goals (G) and tasks (T).
        
        \noindent\textbf{G1 - Immersive workspace:} NDT experts need to inspect rich material data in an immersive environment, taking advantage of location-independent analysis, accurate depth perception, and efficient 3D navigation.\\
        \textbf{G1.1 - Simultaneous analysis of primary and secondary data:} Material characterization requires both primary (structural analysis) and secondary abstract data for the detailed analysis of the attributes' distributions. Therefore, an application must be capable of efficiently supporting both types of analysis.\\
        \textbf{G1.2 - Analysis and comparison for series of rich XCT in-situ tests:} The investigation of changes in the distribution of material attributes over time is critical for experts to understand their behavior under stress from external influences (e.g., forces, etc).\\
        \textbf{T1 - Embodied exploration for spatial volumes:} Experts require embodied navigation and stereoscopic visualization to enable a natural interaction and exploration of inherently spatial material structures (\textbf{G1}).\\
        \textbf{T2 - Overview and comparative characterization of attribute distributions:} Experts require a method to quickly review a large number of attributes across many datasets (\textbf{G1.2}) in order to compare them, filter and identify their value ranges.\\
        \textbf{T3 - Tracking and analyzing temporal changes:} Trends, anomalies, and correlations in the distributions of attributes must be made visible to experts in order to effectively identify the most significant changes over time (\textbf{G1.2}).\\
        \textbf{T4 - Detailed inspection of attribute-specific changes:} Experts require a detailed comparative investigation of changes in the attributes' distributions between time steps and value ranges (\textbf{G1.2}), allowing them to identify anomalies and relate those to the spatial representation (\textbf{G1.1}).

        Due to the limited availability of AR systems for material characterization, the following key research questions were identified to evaluate their usability for analysis of rich material data:\\
        %
        \noindent\textbf{Q1 - Usability and user experience:} How do experts perceive the usability and experience of analyzing rich material data in an AR environment? Specifically, how intuitive and engaging is \framework, in terms of interaction, navigation, and ease of use (\textbf{G1})?\\  
        \textbf{Q2 - Data interpretation and workload:} How do the visualization techniques in \framework influence experts' ability to explore and interpret rich material data (\textbf{G1.1, G1.2})? How does this impact the perceived workload and improves the ability to uncover unseen patterns, understanding correlations and trends, and comprehending attribute distributions?

\section{\framework: Multiview Augmented Reality Visualizations } 
    This section outlines key design considerations for AR visualizations tailored to domain expert tasks. It highlights challenges that shaped \framework and details our novel techniques.

    \subsection{Visualization Design Considerations}
      
        Although AR devices have several advantages, designing interactions and visual metaphors poses significant challenges. See-through AR devices, such as the MH2, typically feature a limited field of view (FoV). This requires the developed visual representations to be very compact to be viewed as a whole. Additionally, a low color fidelity and perspective distortions need to be taken into account \cite{Kraus2020a}. It is also recommended to avoid control components that can cause fatigue due to mid-air interactions \cite{HincapieRamos2014a}.\\
        %
        %
        Other challenges arise from the tasks imposed by domain and experts, as the latter may not always have sufficient information about the target and its location \cite{Brehmer2013}. 
        An understanding of the material's structure or changes over time must be conveyed. Therefore visualization techniques must be able to display a compact representation of several attributes, i.e., distributions, in different units (e.g., angles, areas, volumes). For exploratory data analysis, it is crucial to use statistical parameters as they allow for the comparison of multiple attributes without the need for dimension reduction \cite{Cavallo2019}. Additionally, visualizations should allow for temporal exploration of these attribute distributions to provide experts with a comprehensive insight into the behavior of the material over time. \\
        %
        %
        Although we try to solve most of these challenges in \framework, in the case of color fidelity we refer to existing solutions from Hincapie-Ramos et al.~\cite{HincapieRamos2014}, as this complex issue is beyond the scope of this paper.
        
    \subsection{Visual Representation} \label{subsec:VisualRep} 
        Taking the aforementioned requirements and challenges into consideration, we reviewed the most common visualization techniques based on recent literature \cite{Saket2019, Rodrigues2019, Blumenschein2020}, assessing their applicability to the tasks and goals identified in \autoref{subsec:Requirements}.
        
        Promising techniques for visualizing information from more than three dimensions and different units include parallel coordinates, radar charts, Sankey diagrams, and glyph representations. 
        Some techniques, such as parallel coordinates or radar charts \cite{Waldner2019}, can rapidly become challenging to comprehend due to the abundance of data items and correlations depending on the ordering of the axes (\textbf{T3}). Techniques featuring small multiples or stacks, e.g., histograms or violin plots, are constrained by the available screen space, which in turn results in more difficult juxtaposed comparisons (\textbf{T2}) \cite{LYi2021}. In addition, displaying too many attributes with a high number of data points in a small chart can result in cluttered diagrams and diminish comprehensibility (\textbf{T3}). 
        Area plots may overlap for small changes and perform poor for comparisons \cite{Rodrigues2019}. Box plots may not show the distribution shape effectively \cite{Rodrigues2019, Blumenschein2020}, and histograms can be misinterpreted if the binning is not optimal \cite{Sahann2021, Heim2024}. 
        After evaluating many visualization techniques, we selected glyphs as central idiom for the design space in our study. The glyph's geometric shape resembles bars, encoding relevant statistical measures as their channels. This will flatten the learning curve for users since box plots and histograms composed of bars are familiar graphical methods and commonly used by material experts\cite{Seo2005, Saket2019, Rodrigues2019, Blumenschein2020, Heim2024} (\textbf{T2-T4}). Using simple shapes also eliminates potential misinterpretations caused by perspective distortions.
        
        We combine this familiar geometric form with embodied navigation and avoid menus, which would break immersion (\textbf{G1}). \framework addresses a shortcoming of current visualization systems, which often fail to generate task-specific visual representations \cite{Aigner2011}. Our \firstVis create a natural data exploration workflow for material experts, seamlessly transitioning between visualizations as the user interacts with their components and changes viewpoint (\textbf{T1}). The resulting visualization techniques and the spatial representation are described in detail in the following section.
                   

    \subsection{Visualization and Interaction Techniques} \label{sec:MainMethodology}
        Together with material experts, we have developed the following techniques, which offer a powerful workflow for the analysis of material data in \framework. The techniques are demonstrated through images created with a MH2 in an office environment.
        \subsubsection{Immersive Workspace}

            \framework runs without a workstation on the MH2. Primary data is loaded via a file dialog and rendered directly from the voxel data using DVR, maximum intensity projection, or isosurface rendering (see \autoref{fig:teaser}A). Alternatively, renderings are done using model-based representations, e.g., modeling fibers as cylinders from their radii, start and end points given in the secondary data (\textbf{G1.1}). This approach reduces computational demands and ensures smooth exploration of large datasets. When multiple datasets are loaded, they are arranged in a grid-like structure floating in the AR environment and facilitating efficient comparisons.
            The experts can interact, via hand gestures, with the volumes to adjust their size, position, and orientation for precise analysis (\textbf{G1}) as required for task (\textbf{T1}). MARV allows users to quickly modify visualizations by interacting with them and changing the angle from which they view the charts. We use the ColorBrewer tool~\cite{Harrower2003} to choose the color blind-friendly colors for our visualizations.
            
            
            %

        \subsubsection{\firstVis}
            
            \textbf{Goal:} The main visualization technique of \framework is called Multidimensional Distribution Glyphs (\firstVis). The chart is designed to compactly represent the statistical characteristics of multiple attributes across multiple time steps (supporting task \textbf{T2}). For this reason, we combine the benefits of various representations and metrics, as suggested by Blumenschein et al. \cite{Blumenschein2020}.

            \textbf{Creation:} After loading the datasets, \firstVis serve as an initial representation for analysis and enable the interaction with further visualization techniques. For each loaded dataset, an \firstVis chart is placed in the environment and a 4D \firstVis representation is created that shows all datasets, i.e., time steps, within arms reach and at the user's eye level.
            
\begin{figure}[tbh]
              \centering
              \includegraphics[width=.9\linewidth]{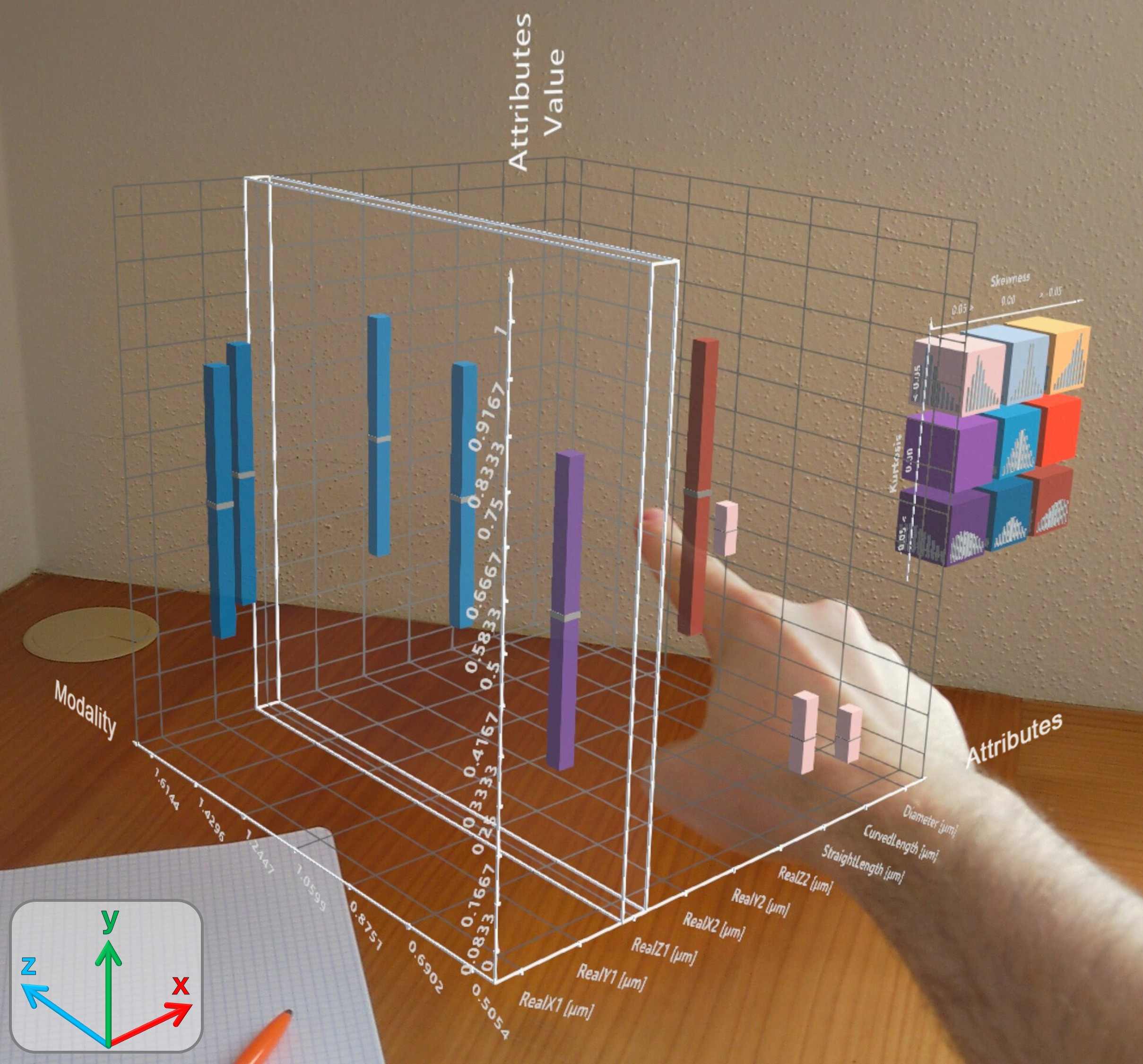}
              \caption{\label{fig:MDDGlyph}
                       \firstVis chart for one dataset: The z-axis is ordered by the distribution's modality. The user points to an attribute on the x-axis to make a selection, indicated by a semi-transparent box with white outline. The \colorWidgetAcro is visible on the right. The coordinate system on the left is added for illustration purposes.}
            \end{figure}

            \textbf{Representation:} The \firstVis chart depicts a three-dimensional coordinate system with the y-axis (\textit{"Attributes Value"}, normalized to $[0,1]$) vertically oriented (see \autoref{fig:MDDGlyph}). The x-axis (\textit{"Attributes"}) displays the attributes of the secondary data. This arrangement facilitates a comparison of values in a single plane. It is formed by the x- and y-axes avoiding perspective distortion and allowing a simultaneous comparison of aggregated attribute values, omitting unit information. Each attribute is represented by a glyph, i.e., a bar, with equal width and depth to avoid perspective distortion. Similar to a box plot, the glyph's center is the distribution's median value and its height is the interquartile range, making it less susceptible to the influence of outliers. The z-axis serves different purposes depending on the task. When analyzing a single dataset, the z-axis (\textit{"Modality"}) displays the modality (shape information) of the distribution. Modality is determined by dividing a distribution into equal-sized bins, summing absolute differences between neighboring bin frequencies, and iterating with varying bin numbers. The maximum sum estimates the number of peaks \cite{Modality}. 
            An ascending ordering according to the number of peaks on the z-axis has the advantage that only distributions of similar shape and thus modality (uniform, unimodal, bimodal, and multimodal), are compared in the plane formed by the z- and y-axes. If multiple sets of time-varying data are loaded, the various time steps are arranged along the z-axis now labeled \textit{"Datasets"}, as can be seen in \autoref{fig:teaser}B. The novel \colorWidget widget, visible in \autoref{fig:teaser}B, addresses some limitations of conventional approaches, such as screen space issues mentioned in \autoref{subsec:VisualRep}. It includes symmetry information of the given distributions by interactively coloring the glyphs by their deviation from a normal distribution~\cite{Seo2005}.
            
            \textbf{Interaction:} Interaction with the \firstVis chart is done through hand gestures, similar as with the spatial representations. A gray cube serves as a handle for spatial manipulation. By changing the orientation of the chart and looking through a different plane formed by two axes, it is possible to observe different aspects of the distributions, such as modality or variability. The view-dependent design of the chart allows experts to generate entirely different representations (see \autoref{subsubsec:secondVis}) based on their viewing direction. This provides users with a flexible and intuitive interaction to explore their data.

            %
            
            %

\begin{figure}[tbh]
              \centering
              \includegraphics[width=.9\linewidth]{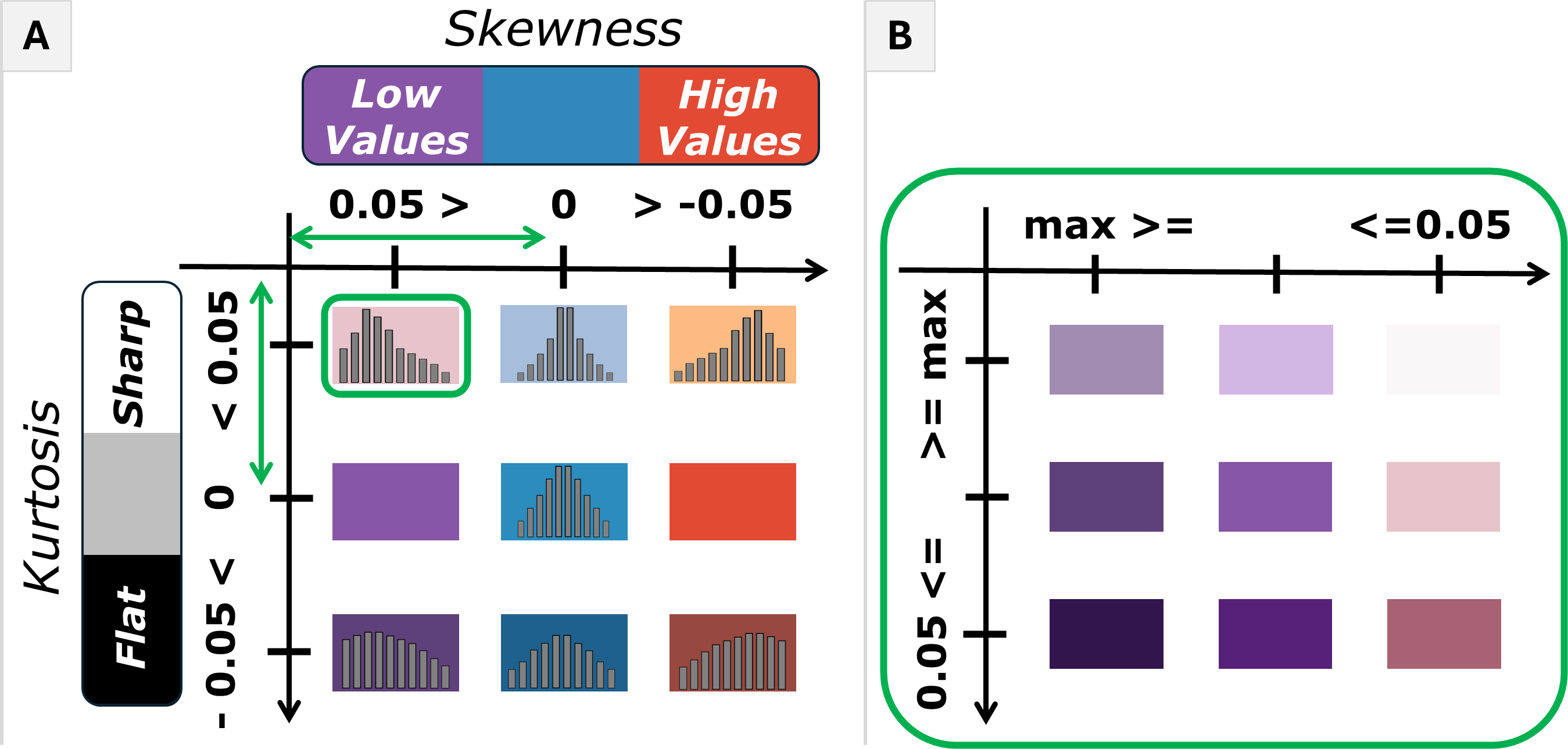}
              \caption{\label{fig:colourScheme}
                       (A) Categorical View: The \colorWidgetAcro features a bivariate color scheme in a $(3\times3)$ matrix that shows the deviation from a normal distribution using kurtosis and skewness.
                       (B) Detailed View: Each matrix cell has a corresponding sub-color space that displays a sequential color scheme, enabling detailed explorations.}
            \end{figure}

            \textbf{\colorWidget (\colorWidgetAcro):} The \colorWidgetAcro (see \autoref{fig:MDDGlyph}) is designed as a bivariate color scheme \cite{Correll2018, Franconeri_2021} that visualizes the deviation from a normal distribution based on kurtosis and skewness. \autoref{fig:colourScheme} depicts the color scheme of the \colorWidgetAcro, called Categorical View, which is arranged in a ($3\times3$) matrix and made up of three individual sequential color schemes (purple, blue, and red). Colors also need to work effectively in an AR environment with changing backgrounds. This led to the rejection of the color schemes \cite{Liu2018} Viridis and Magma due to insufficient luminance against the real background. Black and white were excluded for visualization as they cannot be properly reproduced in AR. The horizontal axis represents the skewness, defined by three categorical colors, indicating whether small or large values predominate. The vertical axis represents the kurtosis, which divides each categorical color into two additional sequential colors, ranging from low to high in luminosity. The kurtosis values indicate the sharpness of the peaks in the distributions. This means that a normal distribution with zero kurtosis and zero skewness forms the center of the matrix.
            
            \textbf{Interaction with the \colorWidget:} Interaction with the \colorWidgetAcro is possible by touching one of the cells in the $(3\times3)$ matrix. In the Categorical View, the user can select further ranges of skewness and kurtosis values. The selected range is then divided into another $(3\times3)$ matrix, allowing the user to zoom in on the desired range of values, creating a new sub-color space called Detailed View. The color scheme for the Detailed View consists of nine sequential colors based on the categorical color of the selected cell (purple, blue, or red) and is applied to the $(3\times3)$ matrix areas shown in \autoref{fig:colourScheme} on the right. This allows for a more precise categorization of the glyphs and subsequently the shape of the attributes' distribution. Any subsequent selection during the Detailed View of any cell in the matrix will restore the original Categorical View color scheme in the glyphs and \colorWidgetAcro.

            \textbf{Use Case - Overview of a single rich dataset:} As shown in \autoref{fig:MDDGlyph}, the expert has loaded an FRP material dataset. It can be seen that the x, y, and z coordinates of the start and end points of the fibers (e.g., RealX1, RealX2, RealY1, etc.) show a high interquartile range, implied by a big bar height. The expert can see that the x, and y coordinates of the start and end points are colored blue and thus are normally distributed. The deviation in length and diameter of the last two bars is smaller, which is indicated by the small bar height. In addition, they feature a strong peak at lower values indicated by a pink color of the bars. Differently, the z coordinates follow a uniform distribution (low modality value). Using the \firstVis, it can be determined for this material sample that the fibers are very similar in length and diameter (similar bar height and color) and are oriented along a certain manufacturing direction (high modality value for x, and y coordinates). In addition, the uniform distribution of the z coordinates indicates a uniform layering of the fibers, confirming a consistent manufacturing process. A similar analysis is possible after loading multiple datasets, with the 4D \firstVis chart displaying all time steps, visible in \autoref{fig:teaser}B.
            

    \subsubsection{\secondVis} \label{subsubsec:secondVis}
        
        \textbf{Goal:} The \secondVis (\secondVisAcro) enables experts to efficiently track attribute changes across multiple time steps. The goal of the chart is to provide a clear and efficient overview of relevant information for complex time-oriented data, allowing for a quick identification of the most prominent temporal changes and their magnitude. The \secondVisAcro chart reflects the domain goal \textbf{G1.2} and addresses the requirements defined by the experts in task \textbf{T3}. By transitioning between \firstVis and \secondVisAcro, users can intuitively compare the most significant changes between time steps (i.e., datasets) without requiring a new chart.
        
\begin{figure}[tbh]
          \centering
          \includegraphics[width=.9\linewidth]{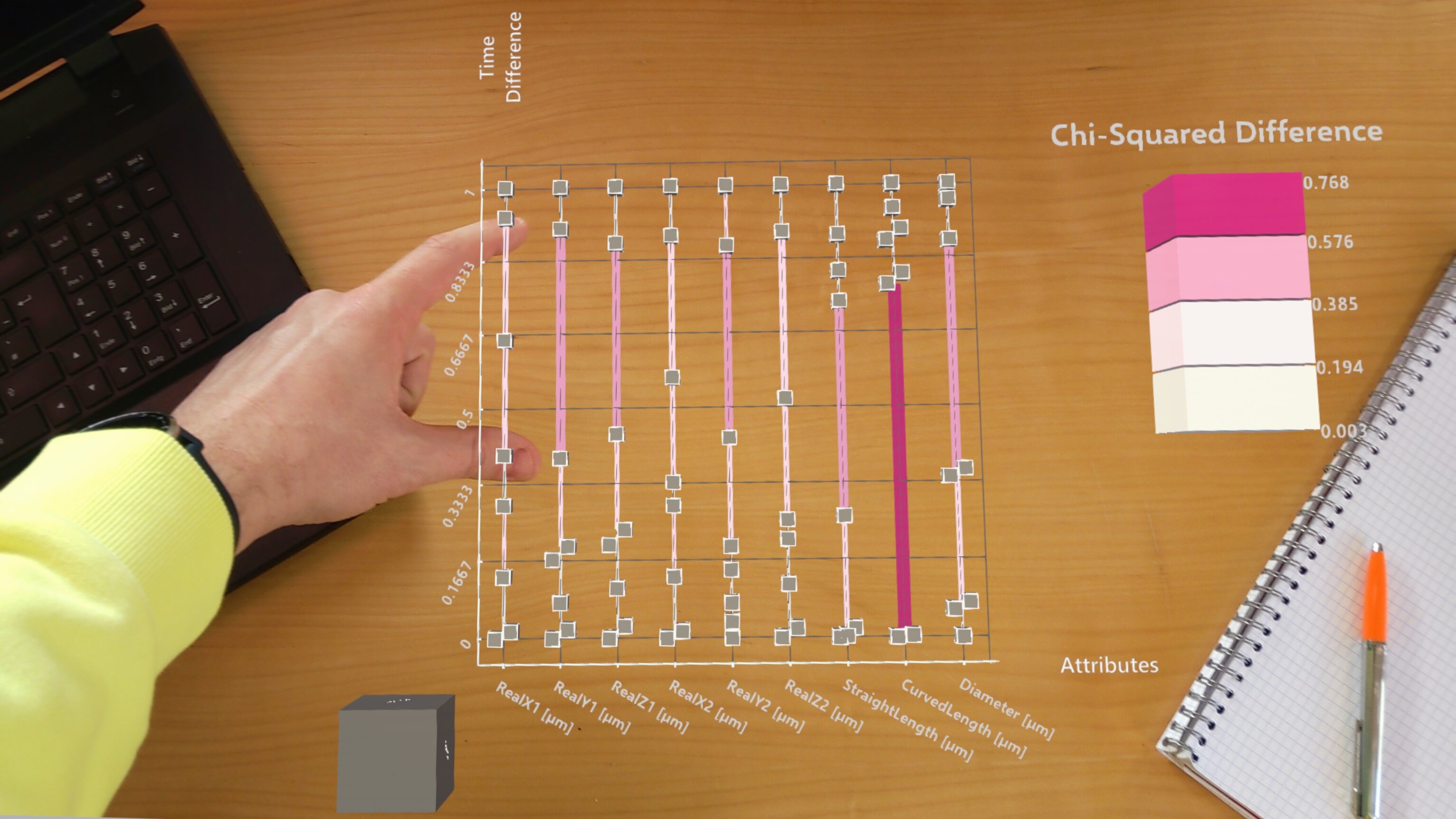}
          \caption{\label{fig:Timescatter}
                   \secondVis: Looking from the top of the \firstVis chart, it is transformed into the \secondVisAcro visualization. \secondVisAcro shows changes in attribute distributions over time. Each of the small cubes represents a point in time. The thickness and color of a line between cubes encode the extent of change.}
        \end{figure}

        \textbf{Transition:} The \secondVisAcro visualization as shown in \autoref{fig:Timescatter} is triggered when interacting with the \firstVis chart by viewing it from the top (or bottom), for example, by placing it on the floor, or rotating it 90 degrees by hand. This view-dependent representation enables quick switching between charts and different analyses based on the user's viewpoint.  
        
        \textbf{Representation:} Rotating the \firstVis transforms it into a 2D chart where the glyphs are abstracted into gray cubes. The x-axis of the \secondVisAcro chart displays the individual attributes of the dataset. The y-axis shows the similarity between attribute distributions at two consecutive points in time, calculated using the Chi-square test \cite{Cha2008}. It is used due to its accuracy in comparing multimodal histograms and lower computational complexity as compared to, e.g., the Earth Mover's Distance \cite{Naik2009, Bazan2019}. The measure is calculated for each consecutive time step of an attribute and then normalized to the interval $[0,1]$. The cubes (datasets) are ordered vertically from bottom to top based on their time step. To improve feature recognition, we encode the differences between time steps redundantly \cite{Heim2024}. This is achieved through the drawn distance between cubes and the color and thickness of lines connecting time steps.
        
        \textbf{Interaction:} Spatial manipulation of the chart itself is possible with a gray cube as handle at the bottom left (visible in \autoref{fig:Timescatter}). Users can freely adjust the position, scale, and rotation of the chart, in line with the requirements for the \secondVisAcro transition.
        
    
        %

        \textbf{Use Case - Temporal changes:} \autoref{fig:Timescatter} presents an overview of the changes over time for a material if additional seven time points of an in-situ test are loaded and the \firstVis is rotated. The \secondVisAcro allows experts to identify that the first and last two time points (cubes) are similar in all attributes, as the cubes are in close proximity to each other. This suggests that the application of force at the beginning and end of the test had minimal influence on the material. Additionally, the attribute "CurvedLength" is of interest to the experts, as the visual encoding hints at a significant shift in the distribution between the second and third load steps, suggesting the fracture of curved fibers. From the sixth time step onward, the experts were able to observe a change in the y and z coordinates of the start and end points of the fibers, indicating a global deformation of the material sample, such as buckling or bending.

    \subsubsection{\thirdVis Visualization}
        
        \textbf{Goal:} The \thirdVis visualization allows experts to analyze the temporal changes of an attribute distribution in detail. This is achieved by calculating a histogram for each time step and stacking the bins vertically. By connecting bins of adjacent time steps with a colored area, trends of individual value ranges can be precisely compared and analyzed. The \thirdVis chart addresses domain goal \textbf{G1.2} and provides experts with the ability to inspect attribute-specific changes in detail, as required by task \textbf{T4}.
        
        \textbf{Transition:} By placing a hand near the \firstVis, the user can select an attribute and all its available time steps, indicated by a semi-transparent selection box with a white outline (visible in \autoref{fig:MDDGlyph}). Dragging the box, i.e., x-axis entry, out of the chart triggers the \thirdVis view.
        
\begin{figure}[tbh]
          \centering
          \includegraphics[width=.95\linewidth]{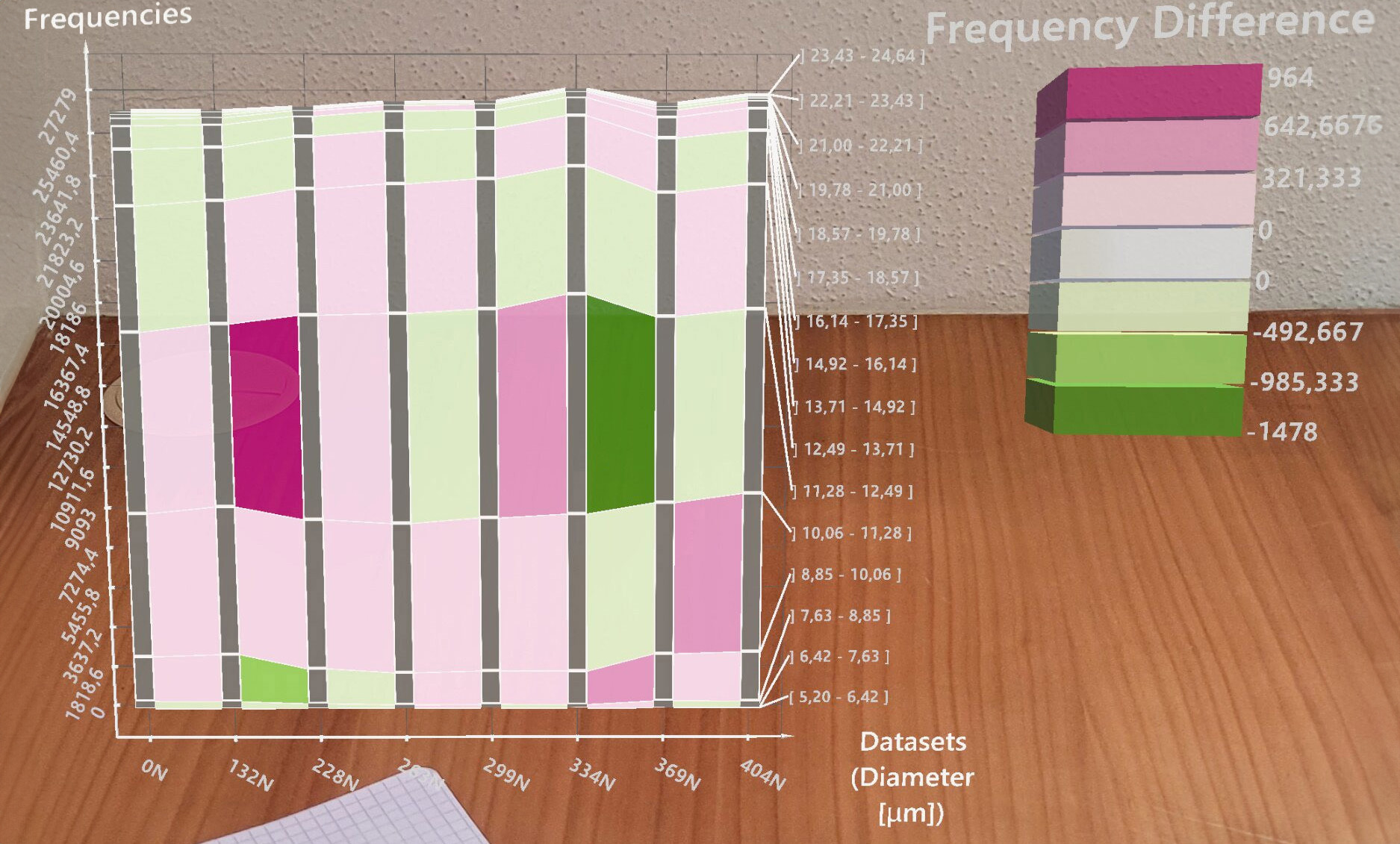}
          \caption{\label{fig:ChronoBins}
                   \thirdVis: Visualize the increase and decrease of data points of a selected attribute as stacked histogram bins. The shape and color of the areas between bins, indicate the number of gained or lost data points, i.e., fibers, between neighboring time steps.}
        \end{figure}

        \textbf{Representation:} The \thirdVis visualization is a 2D chart in which narrow gray rectangular glyphs, i.e., histogram bins, are stacked on top of each other. The individual stacked bars are connected by broad colored areas (see \autoref{fig:ChronoBins}). For each time step (x-axis), a histogram with uniform binning is computed. The number of bins is determined by Sturges' rule for each distribution, and the maximum serves as the bin count for all time steps~\cite{Heim2024}. Sturges' rule offers suitable estimates for over 1000 samples while avoiding the overestimation common in other models~\cite{Sahann2021}. The resulting histograms are shown as narrow gray bars, with heights scaled to the number of elements per bin. 
        The value range of the bins, in the unit of the attribute, is displayed to the right of the final time step at the approximate height of the corresponding bin. This arrangement enables a comparison of bins from adjacent time steps without occlusion as happening in traditional superimposed histograms \cite{Bok2022}. We tested various visual bin widths and selected a suitable size based on expert feedback. To enhance perceiving the trend of changing data points, 
        adjacent time steps are connected by a broad color-coded area indicating the increase/decrease of data points in the value range defined by the bins. 
        To prevent misinterpretations with the colors in the \colorWidgetAcro for the \firstVis chart, we have chosen a different set of colors, indicating increases (magenta), decreases (green) or no changes (gray). The bins were intentionally kept in gray, ensuring that the focus remains on increases and decreases of fiber counts.
        
        \textbf{Interaction:} The chart can be moved, rotated, and resized analogous to the other charts. The \thirdVis visualization can be pushed back into the \firstVis chart (metaphor: "push a book back onto a shelf"), which will remove it. It is also possible to create multiple \thirdVis charts for different attributes at the same time, yielding the opportunity to discover correlations in trends. Clicking on the colored areas triggers a highlighting of the corresponding fibers in the model-based surface representations. Fibers in the value range of the selected bin are highlighted in red and yellow (later time step), in both spatial views. This enables experts to directly observe changes or correlations in the abstract data within the 3D space.
        
    
        %

        %

        \textbf{Use Case - Detailed temporal comparison:} A detailed analysis of the change in fiber diameter over eight time steps is illustrated in \autoref{fig:ChronoBins}. Experts can quickly assess the maximum number of fibers ($27,279$) across all time points and the fiber diameter range ($5.20-24.64~\mu{}m$) from the axis ticks. Moreover, \thirdVis indicate that the majority of fibers have a diameter in the range $8.85-13.71~\mu{}m$ as determined by the visually longest bins. Moreover, the material exhibits an increase in fibers (magenta) of diameters $10.06-11.28~\mu{}m$ between the second time step ($132N$) and the third one ($228N$), accompanied by a loss of fibers (green) of diameters $7.63-8.85~\mu{}m$. Additionally, a contrasting trend is observed between time steps six and seven, suggesting the fracturing of longer fibers. The experts identified an error in the fiber characterization pipeline utilizing our tool: Up to the sixth time step ($334N$), an increase in the total number of segmented fibers can be observed, potentially due to fiber fractures. Afterwards, the fiber count declines, indicating segmentation errors possibly due to failed detection of too small fibers.

\subsection{Framework}
    The \framework prototype can be run either in standalone mode on the MH2 or via remote rendering using a workstation. A Windows 11 notebook with an Intel i9 9800H processor and an NVIDIA GeForce RTX 2080 Max-Q graphics card was used for this purpose. Our tool was developed using the Unity platform (2021.3.14) \cite{Unity} and the MRTK (2.8.3.0) SDK \cite{MRTK}. The source code for \framework is publicly available on GitHub \cite{MARV_GitHub}.
    
\section{User Study} 
    To evaluate the effectiveness of \framework and its novel immersive visualization techniques we conducted a user study on a real-world problem in the field of materials science. The evaluation is guided by the research questions outlined in \autoref{subsec:Requirements}.
    

    \subsection{Participants \& Apparatus}
        We recruited ten participants (four female and six male) aged between 24 and 40 years. All participants have a background in materials science and use various systems to analyze and compare XCT scans. Seven participants describe themselves as experts in materials characterization. All study participants hold a university degree as their highest level of education. Seven of them reported no prior experience with AR. Two participants had moderate experience, while only one person felt somewhat experienced. The vision capabilities of all participants were confirmed to be normal or corrected to normal. The experiment was conducted in a typical office environment where participants could use the application either seated or standing, and they had a free movement area of $2 \times 2$ meters. 

    \subsection{Datasets}
        
        The study utilized XCT scans of a short glass-fiber reinforced polypropylene material as the primary dataset. Interrupted in-situ tensile tests were conducted on the injection-molded material, and it was scanned over several loading steps during these tests (details by Maurer et al.~\cite{Maurer2022}). We deal with primary datasets acquired at eight (temporal) points of the experiment (from zero to 404 Newton loads). Correspondingly, eight sets of secondary data, which contain between \num{26367} and \num{27279} fibers. For the study, we selected the nine most important attributes. As all participants have experience in analyzing fiber datasets, they are asked to perform an exploratory data analysis using our novel AR workflow in \framework.

    \subsection{Procedure}
        The study was conducted in accordance with all the ethical and sanitary guidelines required at the time, and the same procedure was followed for each participant. After completing a demographic questionnaire, they were given a brief 15 minutes introduction to the visualizations and controls of our framework. These were shown on a monitor through remote rendering and using a dataset that was not part of the study.
        The participants then put on the AR HMD and accessed the dataset with the eight load steps. The users were encouraged to test each visualization technique with respective tasks to ensure consistent evaluations. This concerned exploring the differences between attributes (\firstVis), identifying temporal changes (\secondVisAcro), and observing changes in value ranges across time steps (\thirdVis) and their respective spatial visualization. After completion, participants were allowed to explore the other visualizations freely to determine if they provided additional insights or benefits. 
        
        Throughout the study, participants were asked to use a think-aloud protocol to gain qualitative insights into the analysis process and potential usability issues. After each task, the participants were asked to fill in two questionnaires about the visualization techniques employed. We used the NASA Raw Task Load Index questionnaire (RTLX) \cite{Hart2006} to measure the workload and the System Usability Scale (SUS) \cite{Brooke1996} to measure user experience. Even though verbalizing thoughts during the study may have a negative impact on measured workload and usability, we believe that the additional qualitative information is necessary to gain a deeper understanding of user behavior and usability.
        Participants spent approximately 30 minutes in the AR environment. This phase was followed by a semi-structured interview where the participants were inquired regarding their thoughts and comments on the \framework implementation. The study concluded with a short questionnaire where participants rated their AR experience on a 5-point Likert scale and reported any health problems. Each study took about 60 minutes, and all participants successfully completed the trials.
        

\section{Results \& Discussion} 
    To measure the workload associated with our visualization techniques with respect to the tasks to be completed, we use the comprehensive RTLX metric (six questions), where lower values [$0-100$] indicate lower workloads. To facilitate a comparison with other applications, we utilize the reference values outlined in the study by Hertzum \cite{Hertzum2021}, focusing on the most similar technology, i.e., virtual reality. While our evaluation is based on a limited sample of 10 experts, preventing statistical significance, we still consider this comparison important for providing an initial benchmark. To assess the usability of our technologies, we take the commonly used SUS metric (ten questions), where higher scores [$0-100$] indicate better usability. In order to allow for a better analysis of the raw SUS score, we convert the score into grades as indicated by Lewis~\cite{Lewis2018} and Bangor et al.~\cite{Bangor2009}.

    
    \begin{figure*}[htb]
      \centering
      \includegraphics[width=.85\linewidth]{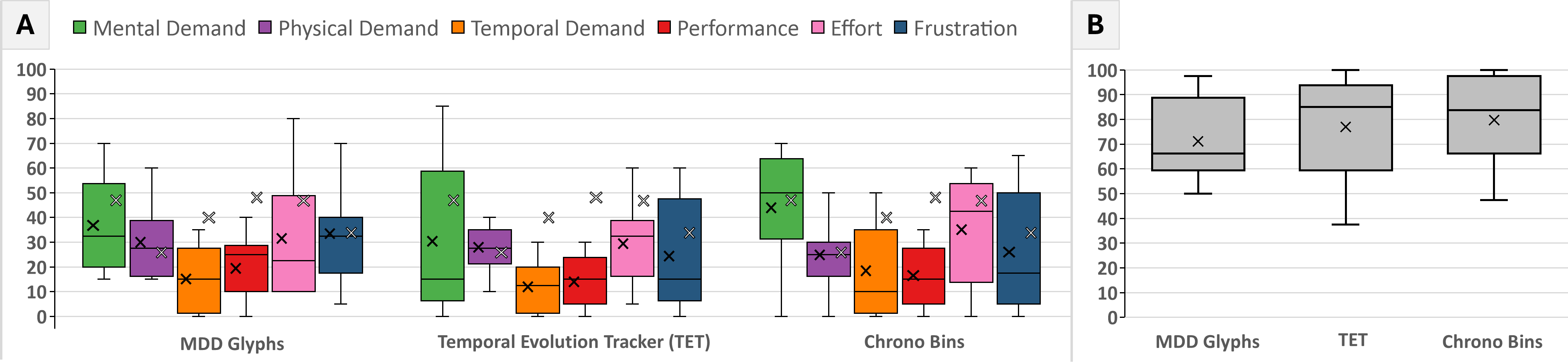}
      \caption{\label{fig:EvaluationResults}
               Study results: Each participant’s responses on the (A) RTLX evaluation and (B) SUS evaluation. For RTLX, lower values are better; for SUS, higher is better. Lines indicate the median, black crosses the mean, and white crosses the mean of the reference.}
    \end{figure*}
        
    \subsection{Evaluation Results}    
        The results of our study indicating that the participants had a positive experience with our immersive framework \framework.
        The RTLX results are shown in \autoref{fig:EvaluationResults}A. In terms of Temporal Demand, Effort, and Performance, all three techniques are significantly below the reference average, indicating that users perceive them as efficient and requiring minimal time investment. While all techniques exhibit lower Mental Demand, compared to the reference values, the \thirdVis chart exhibits the highest value, potentially due to the vast number of visualization tasks it addresses and the associated learning curve. The \firstVis chart has the highest Frustration value, which may be attributed to its role as a primary interaction tool, its novel immersive presentation, and sophisticated color scheme.

        As can be seen in \autoref{fig:EvaluationResults}B, SUS revealed the best usability for \thirdVis (\num{79.75} = A-), followed by \secondVisAcro (\num{77.00} = B) and by a small margin \firstVis (\num{71.25} = C+). Comparing \framework's scores to those of a not acceptable product ($<\num{51.7}$) \cite{Bangor2009}, there are strong indications that immersive analysis is usable for real world industrial applications. 
        %
        The general questionnaire showed a positive attitude to working with AR, indicated by the average response of Somewhat Agree (\num{4.3}). Participants generally agreed that AR could support their work and reported minimal health issues. Detailed results about the evaluation are provided in the supplemental material.
        

    \subsection{Feedback \& Insights}   
        Based on the think-aloud protocol and the interviews, we were able to identify the concerns and suggestions raised by the participants.
        
        Participants liked the compact visualization technique and the view-dependent design of the \firstVis chart (\textit{"I like not having to use a menu [for switching between charts]"}, \textit{"It's great that I don't have to view lots of charts at once, because I can just look at the first chart from above, quickly check for time changes, and then get back to the full view"}). Users also mentioned that the holistic overview helps during analysis as \textit{"It is convenient to see all the characteristics of the fibers at once"}. Participants also liked the fact that they did not need a controller, mouse, or keyboard. \textit{"It [the interaction] feels much more natural and is more fun than with our tools. And I can do it [the analysis] anywhere,"} they said. \\
        The system as a whole was also well received. During the interview, it was mentioned that \textit{"I can simply add new samples and organize their charts and representations in the room. So I always know in which area I have which load steps [time steps]"}. The \thirdVis display in particular received a lot of positive feedback. The participants liked the intuitive extraction of attributes, the combination of trend analysis, and investigation of changes in the spatial representation. They mentioned \textit{"It was great to simply grab an axis and place it next to each other [in the room] and compare them"} and \textit{"... simply highlighting values of a time step in the volume is very helpful for our work"}. We have also received feedback that \textit{"\framework can be great for showing novices and stakeholders the impact of certain characteristics directly in the spatial representation"}.

        The think-aloud protocol revealed some issues requiring further investigation. Users sometimes had interaction problems due to tracking errors for interactions near their body. This was evident from verbal comments like \textit{"the gray cube handle is too small to touch"} and for the \thirdVis \textit{"... now I've touched the wrong area [between the bins]"}. This could be improved by enlarging the charts, but led to the comment to add a coordinate system in the spatial view to give a \textit{"sense of scale"}. Two participants also noted that the gray color and small size of the bins for the Chrono Bins were difficult to differentiate from the background, but as a solution, they increased the size of the chart. Feedback in the interview revealed a steeper initial learning curve for some users - \textit{"It is quite difficult at first to know how to read the chart [\firstVis] and what gestures you can perform on it"}. Further feedback was provided additional utility features. This includes automatic alignment of charts on desks and sorting of spatial representations by time step or name instead of loading order. To avoid clutter, we intentionally excluded two histogram images in the \colorWidgetAcro, relying on the remaining visuals to convey the distribution changes. However, this caused confusion for one participant, who questioned, “Why are two histograms missing from the selection field [\colorWidgetAcro]?” One interesting comment was \textit{"I am used to looking at 2D sectional views, so it would be great to see these alongside the volume"}. 
        

    \subsection{Discussion \& Limitations}

        The user study offered valuable insights into \framework's effectiveness, highlighting both strengths and limitations. Positive SUS scores and feedback demonstrated high usability and a positive user experience. Experts valued the location-independent analysis, intuitive embodied navigation, and menu-less switching between visualization techniques. This creates an efficient and engaging workflow, addressing \textbf{Q1}. Low RTLX scores, combined with think-aloud protocol and interview insights, showed that \framework’s immersive visualization techniques enabled material analysis, addressing \textbf{Q2}. The framework supported experts in uncovering patterns, correlations, and trends in time-dependent attribute distributions while reducing cognitive workload. Compact visualization arrangements and the ability to highlight temporal changes further improved understanding of multidimensional data, achieving domain goals \textbf{G1.1} and \textbf{G1.2}. Overall, our system enables experts to analyze time-varying XCT data of materials effectively through immersive visualization and interaction, fulfilling domain goal \textbf{G1}.      
            
        We have discovered limitations in \framework that go beyond those related to usability. These include the restriction to primary data of limited size (several hundred Megabytes) necessary to enable interactive analysis. Environmental conditions may present a limitation due to bright backgrounds and limited color fidelity. Additional visualization techniques could improve the analysis even further, e.g., charts in IATK~\cite{Cordeil2019}, and integrating additional interactions between 2D and 3D visualizations, as in STREAM~\cite{Hubenschmid2021}. Adaptions of spatial renderings (semi-transparent rendering of unchanged fibers) and overlays of the XCT data on real material samples were further requests by experts. The display of the number of fibers within the bins of the \thirdVis chart represents a further potential improvement, which might be addressed by a zoom-in interaction. 
        Another promising approach involves complementary interfaces, combining \framework with tools like open\_iA~\cite{Froehler_2019}, which is likely beneficial to experts as hinted in literature~\cite{Gall2022, Wang2022a}. In terms of the limitations of our user study, the number of material experts who volunteered to participate was relatively low. It would be advantageous to conduct a larger study with a greater number of participants. 

\section{Conclusion \& Future Work} 

    We presented a novel immersive AR framework, \framework, designed to support materials science experts in analyzing rich material data obtained from composites. We have presented three new visualization techniques, each revealing different aspects of the data. 
    We conducted a qualitative study with materials experts using standardized questionnaires to evaluate the framework. It confirmed that analyzing distributions in AR environments is feasible with low workload. The feedback collected during the interviews provided valuable information on how \framework and the overall IA analysis process can be improved. Since our techniques primarily operate on distributions of attributes, their applicability extends far beyond the field of materials analysis, e.g., to climate science, bioinformatics, economics, or social sciences. \\  
    Although we developed \framework using familiar idioms, we discovered expert interest in traditional methods such as 2D slice images, highlighting the need for a gradual transition between known domain tools and new immersive applications. Future work should aim at exploring these transitions, taking advantage of the added spatial interaction. Finally, we would like to conduct a quantitative study comparing our visualization techniques with appropriate traditional 2D and 3D diagrams to determine differences in completion time and accuracy. \\
    %
    Overall, our research highlights the potential of \framework to leverage the users' spatial cognition for more effective visualization of abstract and spatial data, providing new insights into material characterization and quality control.



\section*{Conflicts of Interest} 
The authors declare no conflict of interest.

\section*{Data Availability Statement}
The data that support the findings of this study are available on request from the corresponding author. The data are not publicly available due to privacy or ethical restrictions.

\bibliographystyle{eg-alpha-doi}
\bibliography{mainRef}



\end{document}